\documentclass[review]{elsarticle}

\usepackage{lineno,hyperref}
\usepackage{amsmath}

\journal{Journal of \LaTeX\ Templates}

\begin{document}

\begin{frontmatter}

\title{A Study of NaI(Tl) crystal Encapsulation using Organic scintillators for the Dark Matter Search.}

\author[a]{J.~Y.~Lee}
\cortext[mycorrespondingauthor]{Corresponding authors}
\ead{jylee8875@gmail.com}

\author[b]{G.~Adhikari}

\author[c]{C.~Ha}
\ead{changhyon.ha@gmail.com}

\author[a]{H.~J.~Kim}
\author[c]{N.~Y.~Kim}
\author[d]{S.~K.~Kim}
\author[b,c]{Y.~D.~Kim}
\author[c]{H.~S.~Lee}

\address[a]{Department of Physics, Kyungpook National University,\\Daegu 41566, Republic of Korea}
\address[b]{Department of Physics, Sejong University,\\Seoul 05006, Republic of Korea}
\address[c]{Center for Underground Physics, Institute for Basic Science (IBS),\\Daejeon 34126, Republic of Korea}
\address[d]{Department of Physics and Astronomy, Seoul National University,\\Seoul 08826, Republic of Korea}

\begin{abstract}
   Scintillating NaI(Tl) crystals are used for various rare decay experiments, such as dark matter searches.
The hygroscopicity of NaI(Tl) crystal makes the construction of crystal detectors in these experiments challenging and requires a tight encapsulation to prevent from air contact.
More importantly, in a low radioactivity measurement, identification of external radiations and surface contamination is crucial to characterize the origin of total crystal radioactivities.
Studies for NaI(Tl) crystal encapsulation with active organic scintillator vetoes have been performed to mitigate the above-mentioned issues simultaneously.
A bare crystal is directly coupled with liquid and plastic scintillators to tag external radiations that penetrate from the outer part of the crystal.
We report the pulse shape discrimination for organic scintillator pulses from those of the crystal scintillator in a single detector setup which makes external gammas identifiable and long-term stability tests of the detector setup.
\end{abstract}

\begin{keyword}
  NaI(Tl), organic scintillator, dark matter, phoswich
\end{keyword}

\end{frontmatter}

\section{Introduction}

    Since first developing Tl-doped NaI scintillator for radioactive detection~\cite{Hofstadter},
it has been employed across a wide range of applications
from environmental monitoring to rare nuclear event searches.
Their relatively high light yield~\cite{12684} and wide dynamic range make them more attractive than any other crystal scintillators.
Continued light yield improvement has made the crystals usable to a low energy dark matter experiment since the early 1900s~\cite{crystals}.
These searches look for the scintillation light induced by nuclear recoils in an energy spectrum when a weakly interacting massive particles (WIMPs) interacts in an array of NaI(Tl) crystal detectors~\cite{dama}.

Typical experiments run an array of segmented detector modules, because
the production of high-quality, large volume crystals that is technically challenging and cost ineffective in general.
Those modules are individually encapsulated under low humidity N$_{2}$ gas environment to protect their surface from direct contact with air.
Even so, many dark matter experiments show remnant $^{210}$Pb contamination due to $^{222}$Rn or other radioactive impurities~\cite{edelweiss,SuperCDMS,LUX}, the most dangerous background sources.

Deposited alpha particles in the crystal would be identified by using pulse shape discrimination(PSD) method~\cite{alphapsdkims}, but escaped from the surface alpha particles could be misclassified as WIMP-nucleon interaction due to the deposited kinetic energy of daughter isotope signal which resembles $\gamma$/$\beta$~\cite{surfacealpha}.
Therefore, it is important to reject surface alpha events for dark matter search experiments which is using PSD for separation.
A concept of phoswich detector is the one solution to separate for the $\gamma$/$\beta$ and alpha by using PSD. The detector is constructed with several scintillators which have different decay time~\cite{phoswich}.
The CRESST-II introduced scintillating housing to detect free alpha particles, and they successfully shifted those background events from the region of interest~\cite{surfacealphapsd}.

This paper reports direct encapsulation of bare NaI(Tl) crystal using an organic scintillator as an active veto has been developed
to resolve both hygroscopic issue and the surface radioactive contamination together,

\section{Experimental setup}

\subsection{NaI(Tl) crystal encapsulation with liquid scintillator}
\begin{figure*}[!htb]
  \begin{center}
      \includegraphics[width=0.9\textwidth]{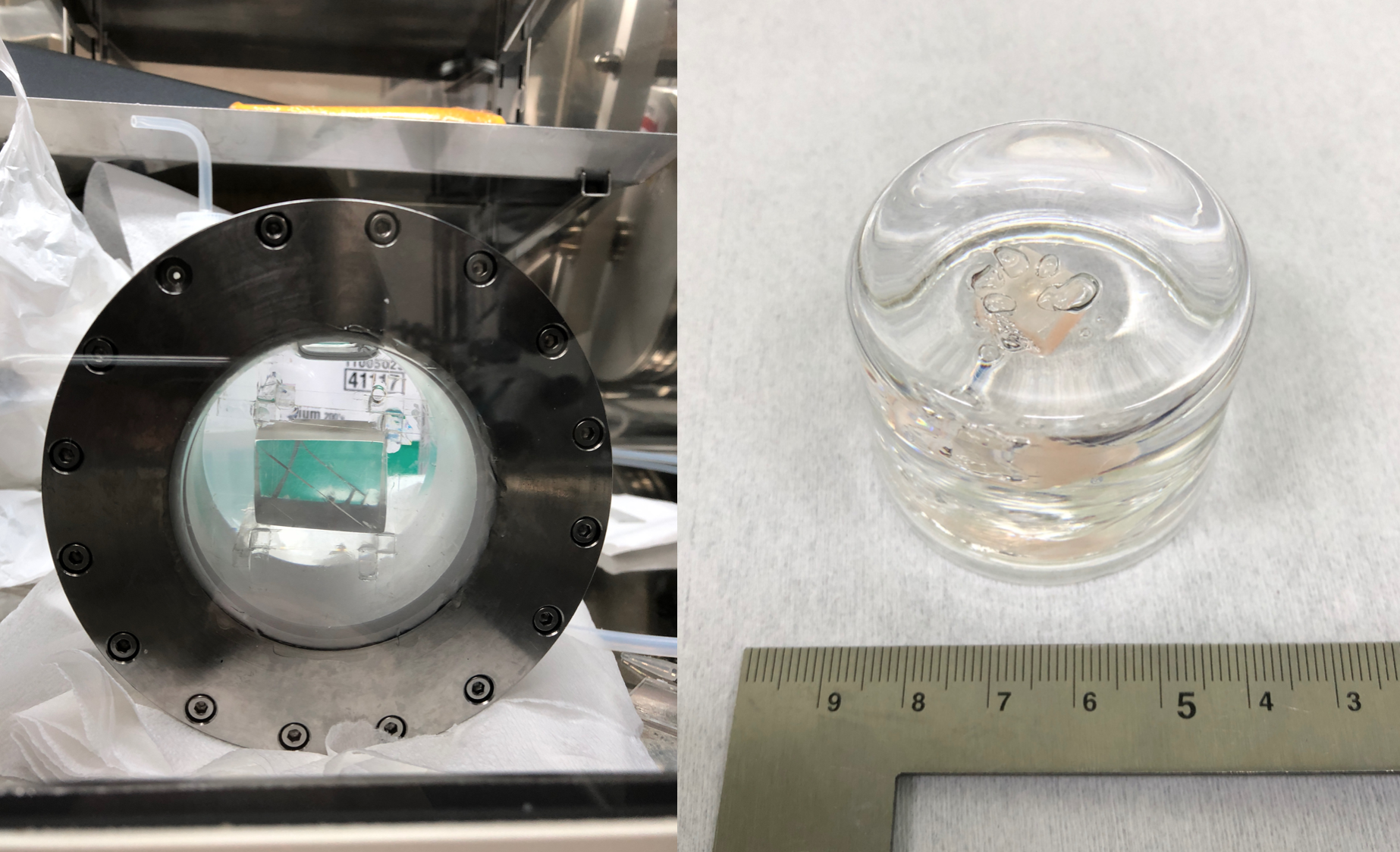}
  \end{center}
  \caption{Proposed NaI(Tl) encased with liquid scintillator (left), and plastic scintillator(right).
  }
  \label{NaI}
\end{figure*}

 Figure~\ref{NaI}~(left) shows the experiment setup, a bare NaI(Tl) crystal is submerged in a liquid scintillator (LS) filled container.
 The crystal is positioned in the center of container in direct contact with LS,
and stabilized with a small acrylic structure.
The cylinderical PTFE container is 50~mm-thick with 80~mm diameter and length.
The container also works as a diffusive light reflector.
Two quartz glasses with stainless steel covers
are used to light-couple and to seal the detector.

Properties of the LS have been widely studied in many experiments
requiring a large volume and/or irregular shape of a detector.
Among a wide range of solvents, we select linear alkyl benzene~(LAB), which has becomes an increasingly popular due to its low-toxicity and cost-effectiveness.
This, LAB-LS, includes a few percents of 2,5-diphenyloxazole dissolved in the solvent for fluorescence and a trace amount of 1,4-bis[2- methylstyryl]benzene~(bis-MSB) to improve the light properties~\cite{prototype}.
The LAB-LS production is performed under controlled environment by supplying high-grade Argon gas
to prevent light output degration caused by oxygen quenching~\cite{neosls}.

The NaI(Tl) crystal is procured from Alpha Spectra Inc.,
who provided similar quality crystals of the COSINE-100 experiment~\cite{cosine}.
Typical low-radioactivity crystals from Alpha Spectra Inc. exhibit 15 photoelectrons per keV light output. Other properties of this scintillator are summarized elsewhere~\cite{cosine,naicrystal1,naicrystal2}.
The particular crystal is rectangular 25~mm~$\times$~24~mm~$\times$~38~mm.

The NaI-LS detector is coupled to the 3-inch photomultiplier tubes(PMTs) made by Hamamatsu(R12669SEL) on each side, and
then it was installed inside low-background shields of the korea invisible mass search~(KIMS) experiment facility in the 700-m-deep underground at the Yangyang Pumped Storage Power Plant~\cite{kimsroom}.
An air-conditioner maintained (23.0$\pm$0.1)~$^{\circ}$C inside the shields throughout this measurement.
Signals from both PMTs are amplified by a factor of 30, and a 400~MSPS~\footnote{magasampling per second} fast analog-to-digital converter(FADC) was used as a digitizer.
We used a disk-shaped $^{60}$Co gamma source ($\sim$1~$\mu$Ci) that emits 1173 and 1332~keV gamma-rays.

\subsection{NaI(Tl) crystal encapsulation with a plastic scintillator}
In this setup, the plastic scintillator (PS) was used as a veto detector as well as encapsulation material.
A bare cubic NaI(Tl) crystal is submerged in the partially polymerized scintillating liquid\footnote{
The product EJ-290 from Eljen Inc. is used.}
which is slowly hardened as a plastic scintillator to encase the bare crystal
as shown in Fig.~\ref{NaI}(right).
We perform the hardening process according to the manual at the beginning, but it is unsuitable for NaI(Tl) encapsulation due to the temperature increment.
We suspect the process causes the bubbles and give a boost to a chemical reaction between plastic and crystal.
Therefore, this casting procedure is carried in a low-humidity metallic glove box with Argon(96\%) and Hydrogen(4\%) gas mixture supplied and a dehumidifier under controlled temperature environment.
The NaI-PS is machined into a cylindrical shape with 50~mm diameter and 40~mm length,
and it is polished to make the surface transparent.
It was then, wrapped with several Tetratex sheet layers for light reflection
and one end of the cylinder was light-coupled with a 2-inch Hamamatsu PMT (H7195).
We used a 500~MSPS FADC digitizer with improved timing resolution and dynamic range.
This measurement was carried out in a ground-floor laboratory.

\section{Methods}

\subsection{Pulse Shape Discrimination}
The pulse shape of the output signal from a scintillator
can be used to identify different radioactive isotope types,
because the interaction of a heavier particle can make cause transition
to more complex molecular or lattice energy states~\cite{naipsd,liquidpsd}.
Typical observables take advantage of the pulse's decay time.

Similarly, in this hybrid scintillator setup,
we identify NaI(Tl)-induced signals from organic scintillator-induced signals
using their intrinsically different decay times.
The decay time of a NaI(Tl) crystal is around 230~ns, whereas organic scintillator has
typically less than 10~ns~\cite{knoll}.

Here, we use the discrimination observable ``meantime''~\cite{kimscsi},
amplitude-weighted average time of a pulse, defined as
\begin{equation}
  \text{Meantime (ns)} = \frac{\sum_{i}^{bins}{q_it_i}}{\sum_{i}^{bins}{q_i}} ,
\end{equation}
where $q_i$ and $t_i$ are $i$-th bin amplitude and its time respectively.
$Bins$ run for 500~ns from the trigger position.
Meantime is strongly correlated with the pulse's decay time,
it presents slightly shorter time than the decay time.
Since it only takes into account the shape, its minimal energy dependence helps
reduce the bias due to event selection.

\subsection{Signal shape of crystal scintillator versus organic scintillator}
\begin{figure*}[!htb]
  \begin{center}
      \includegraphics[width=0.9\textwidth]{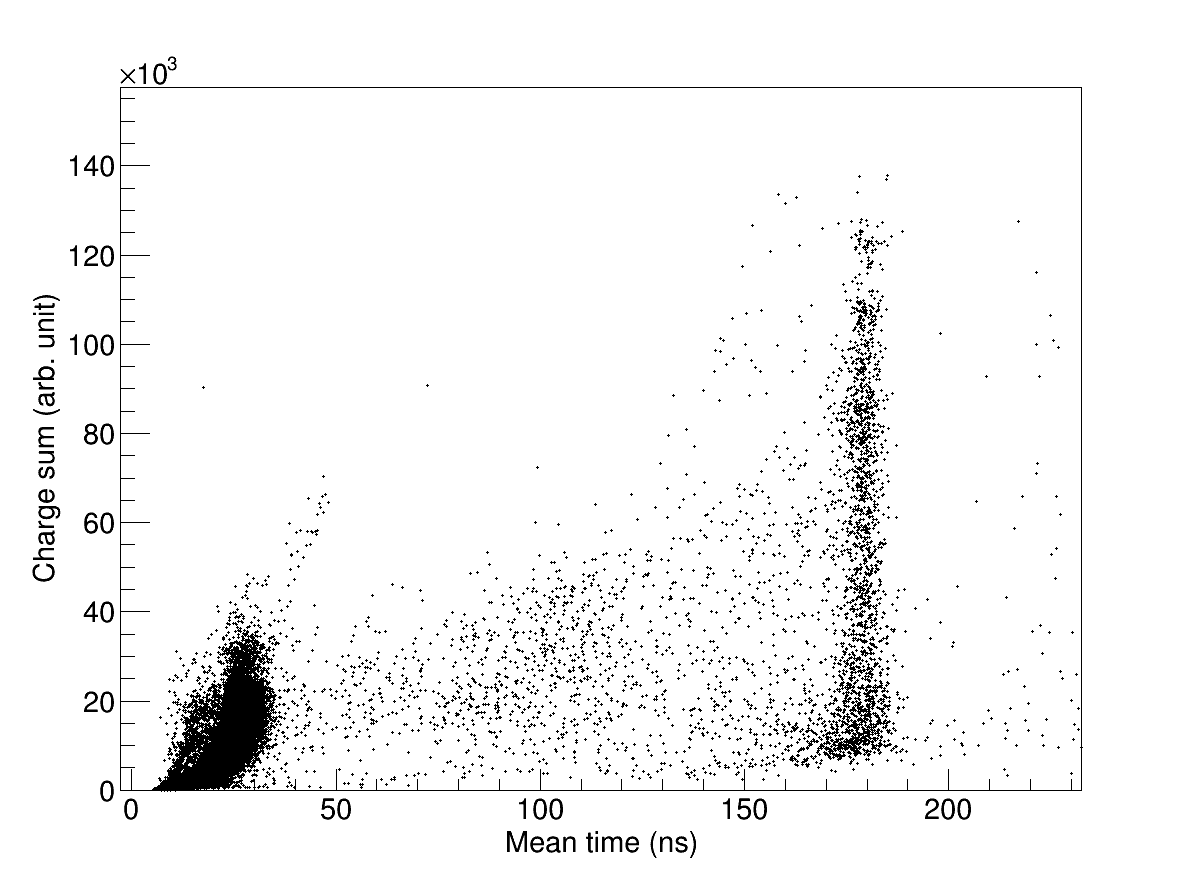}
  \end{center}
  \caption{Mean weighted time distribution of crystal-plastic detector.
    Plastic scintillator and crystal signals are well separated.
  }
  \label{meantimeps}
\end{figure*}

The plastic scintillator used in this experiment is known to have short decay time around $\sim$ 3 ns~\cite{plastic}.
Thus, the maximum signal height from the plastic scintillator shows a relatively higher value than that from the NaI(Tl) crystal scintillator for the same deposited energy.
A wide dynamic range of DAQ is required to measure plastic and NaI(Tl) signals simultaneously.
Figure~\ref{meantimeps} shows the meantime distribution as a function of electron-equivalent energy
that the variable can separate different scintillation signals from the plastic scintillator, crystal scintillator, or both scintillators with.
Two prominent bands at approximately 25 and 180~ns.
Events between those two bands contain merged scintillation signals.
The crystal produces more light from deposited energy than does the plastic scintillator
due to their quenching factor.
Additionally, a full peak of the source is seen in the crystal while that is not well observed in the plastic
due to their different range characteristics.

\begin{figure*}[!htb]
  \begin{center}
      \includegraphics[width=0.9\textwidth]{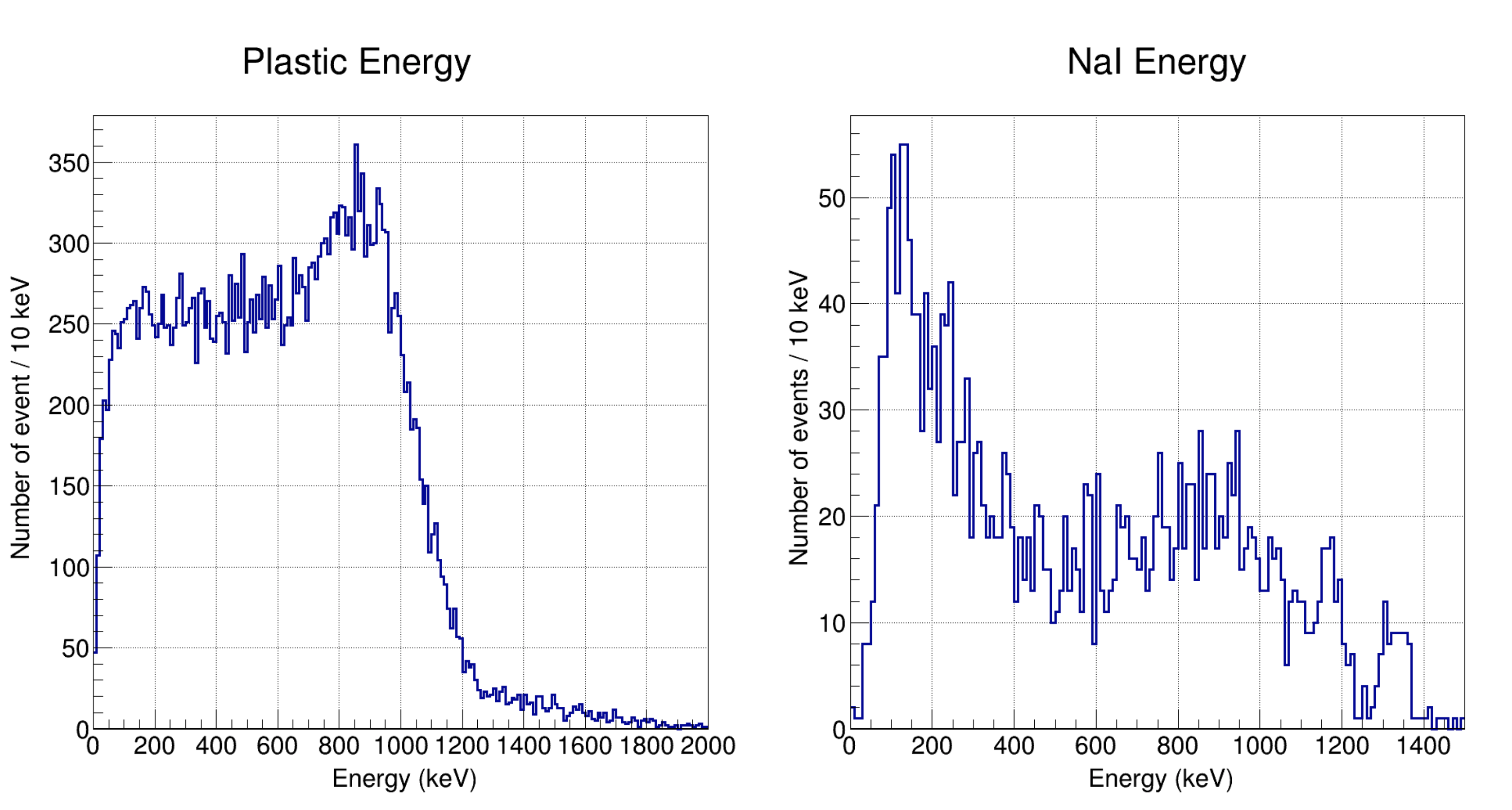}
  \end{center}
  \caption{Plastic and crystal scintillator energy spectra.
    Compton edge fitting is used to calibrate the energy spectra of plastic scintillator,
    whereas the full peaks are used for crystal.
  }
  \label{meantimepsenergy}
\end{figure*}

Figure~\ref{meantimepsenergy} shows the energy spectra for each band
after selecting events within 3$\sigma$ from the mean of each band.
The energy spectra of crystal is calibrated using the known peaks of $^{60}$Co,
whereas the plastic is calibrated by using the Compton edge.

\subsection{Merged signal identification}
\begin{figure}[h]
  \centering
  \begin{tabular}{ccc}
  \includegraphics[width=0.45\textwidth]{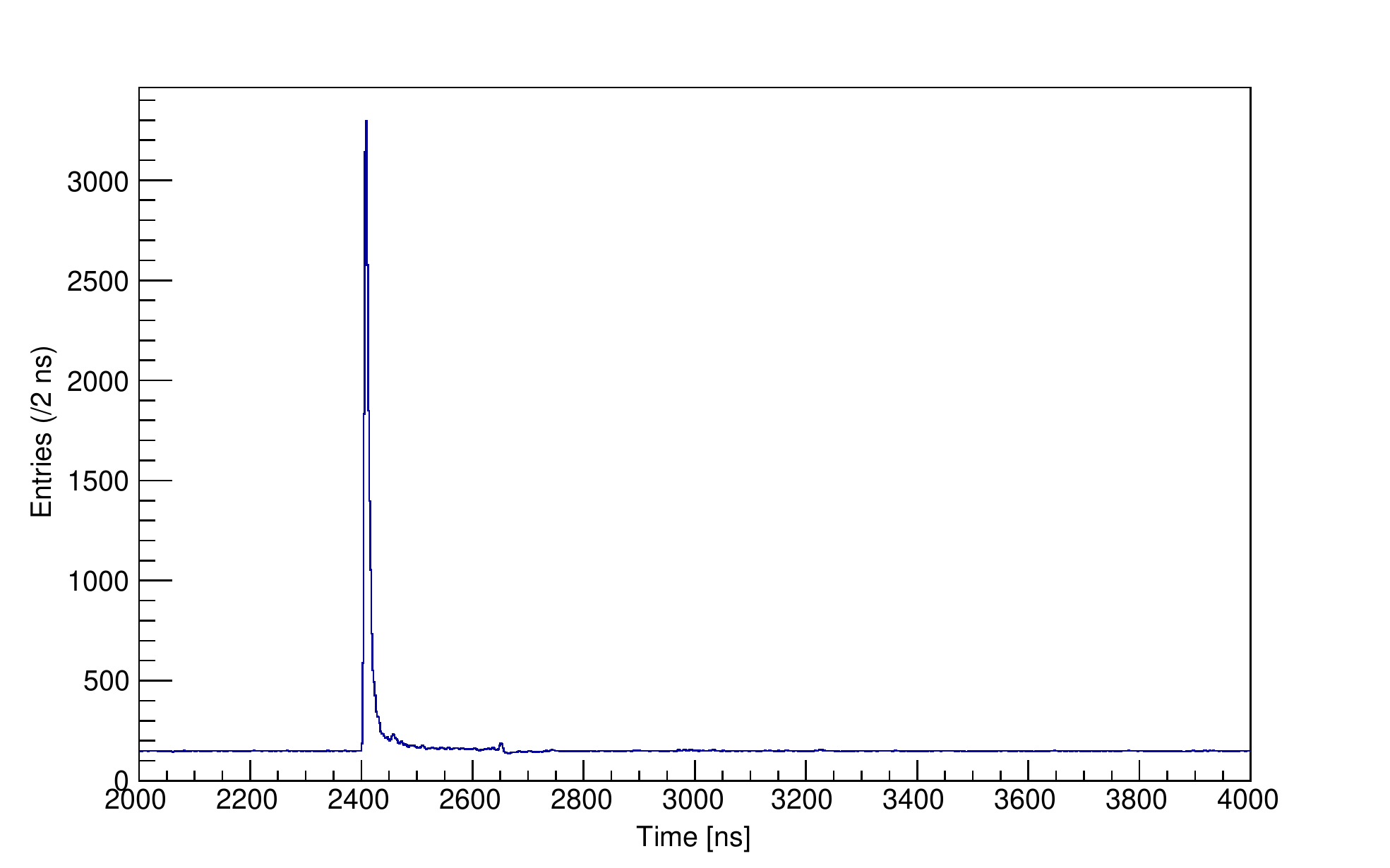} & \includegraphics[width=0.45\textwidth]{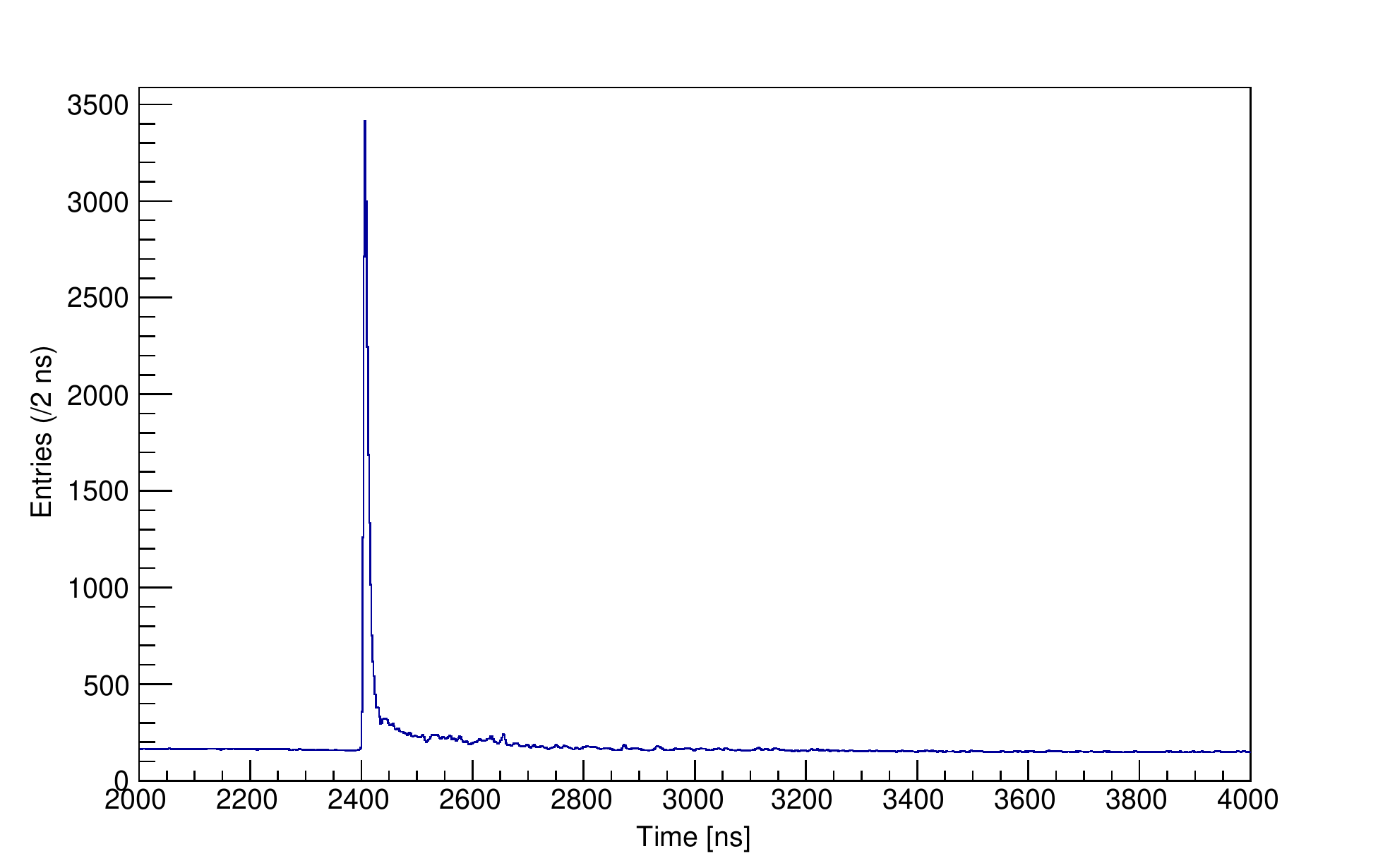} \\
  \includegraphics[width=0.45\textwidth]{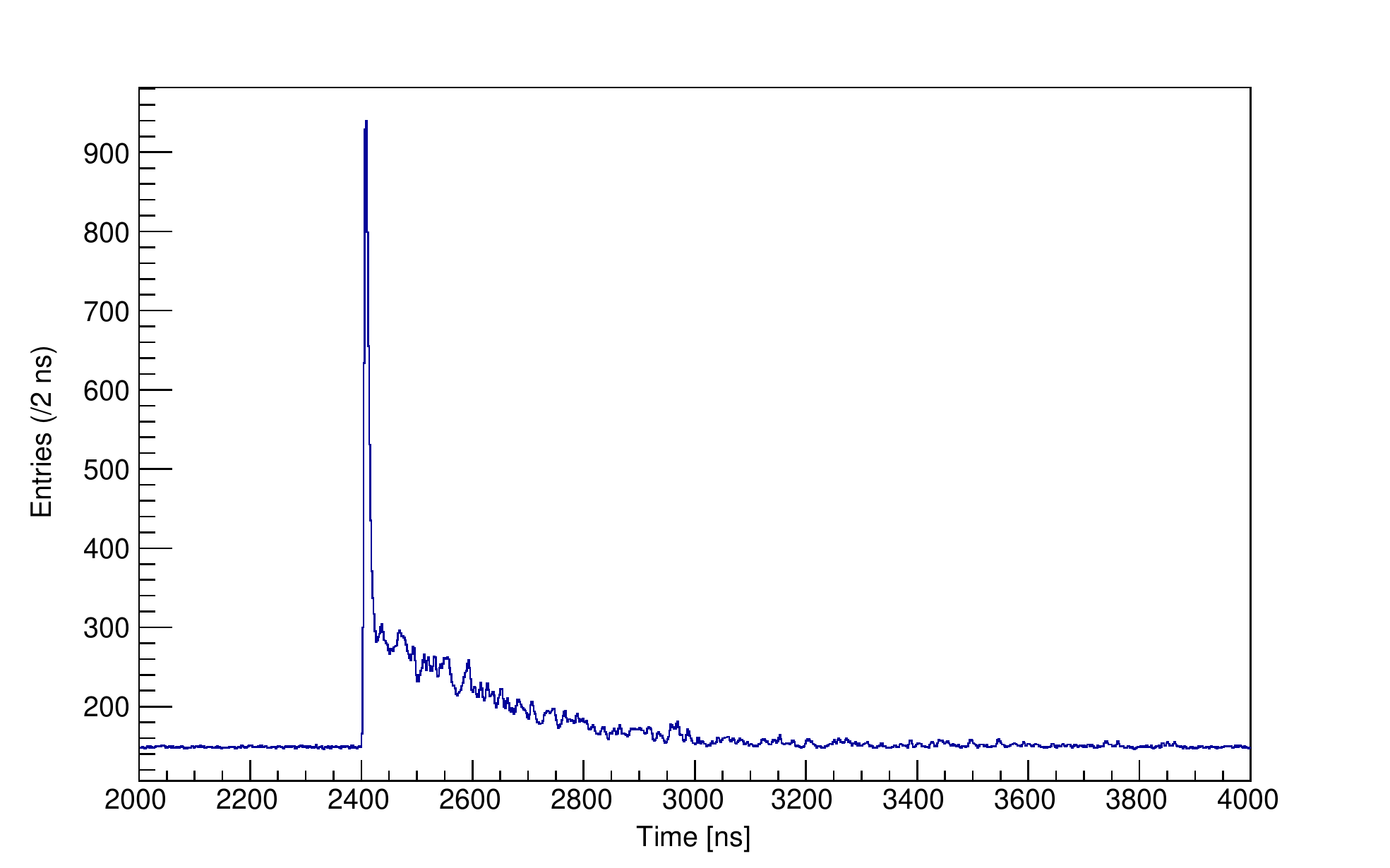} & \includegraphics[width=0.45\textwidth]{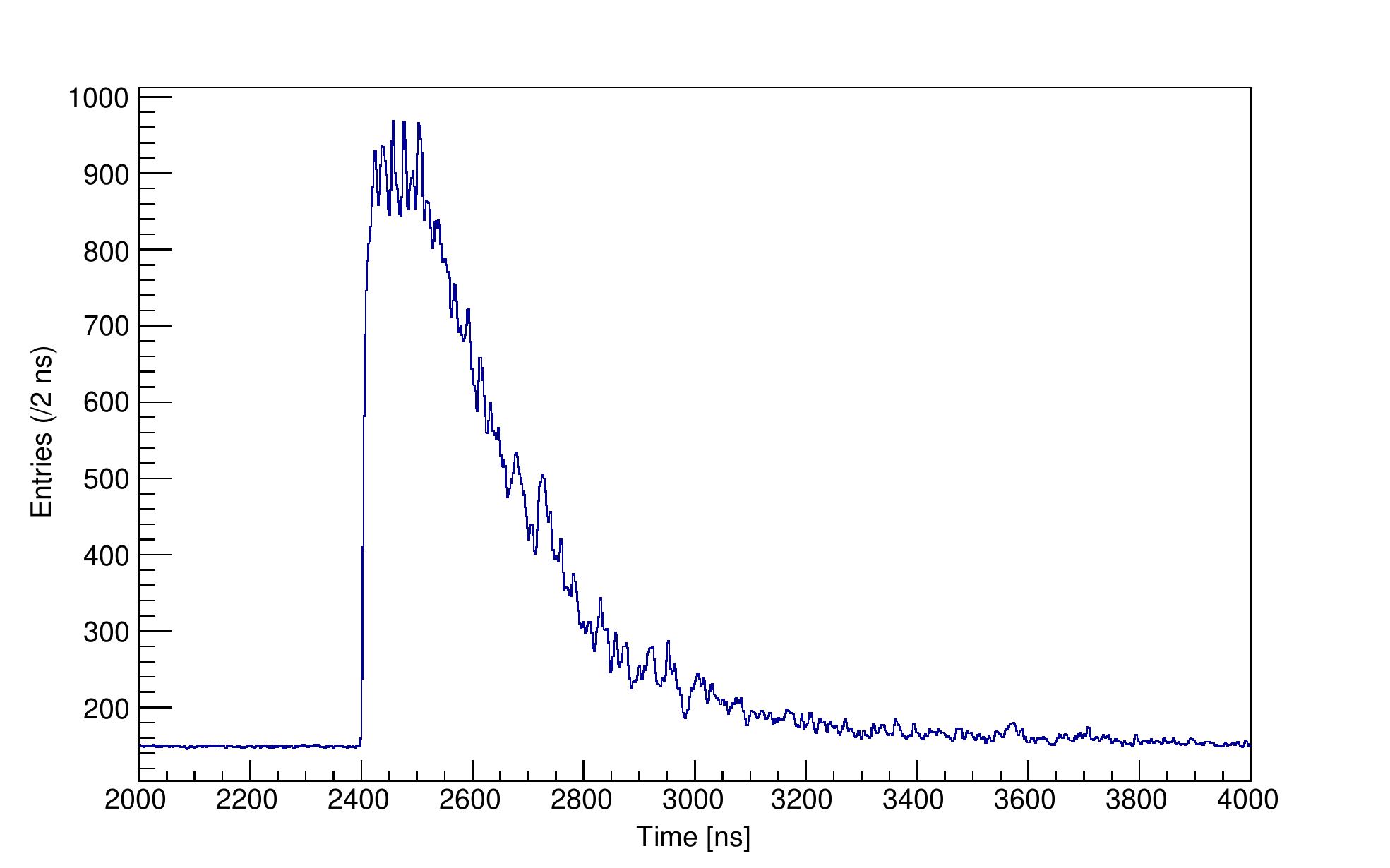} 
\end{tabular}
  \caption{Raw waveforms depending on the signal types.}
  \label{waves}
\end{figure}

Merged signals can occur when the particle interacts with both crystal and plastic.
This signal can be caused by Compton scattering or surface alpha decay process near the boundary,
as shown in Figure~\ref{waves}.
They were selected in the same energy range (200 - 400 keV) but different meantime.
Figure~\ref{waves} (a) and (d) figures are from plastic and crystal events, respectively,
whereas (b) and (c) show mixed signals from both scintillators.

\subsection{Low temperature measurements of NaI-LS}
Both NaI-LS and NaI-PS tests, their light yields show decreasing behavior during the measurements.
We suspected that remnant moisture, moisture diffusion, or unknown chemical reactions would affect the NaI(Tl) surface quality which influences on the light yield.
In this test, the NaI-LS detector was put in the refrigerator to freeze moisture in the LS.

We used a commercial freezer which kept at -35$^{\circ}$C.
The calibration factor and light yield were regularly measured at the Kyungpook National University laboratory
using $^{241}$Am radioactive source, emitting 59.54 keV gamma-rays.
Consequently, NaI(Tl) exhibits 11.3 $\pm$ 1.0 ph/keV light yield.
Considering light yield variation~\cite{lowtempnai}, systematic uncertainties, and PMT effect,
this result is comparable with other crystals from the same company.
Light yield of NaI-LS detector at the room temperature is decreasing over time,
whereas light yield was constant at the low temperature environment 121 days.

\begin{figure*}[!htb]
  \begin{center}
      \includegraphics[width=0.9\textwidth]{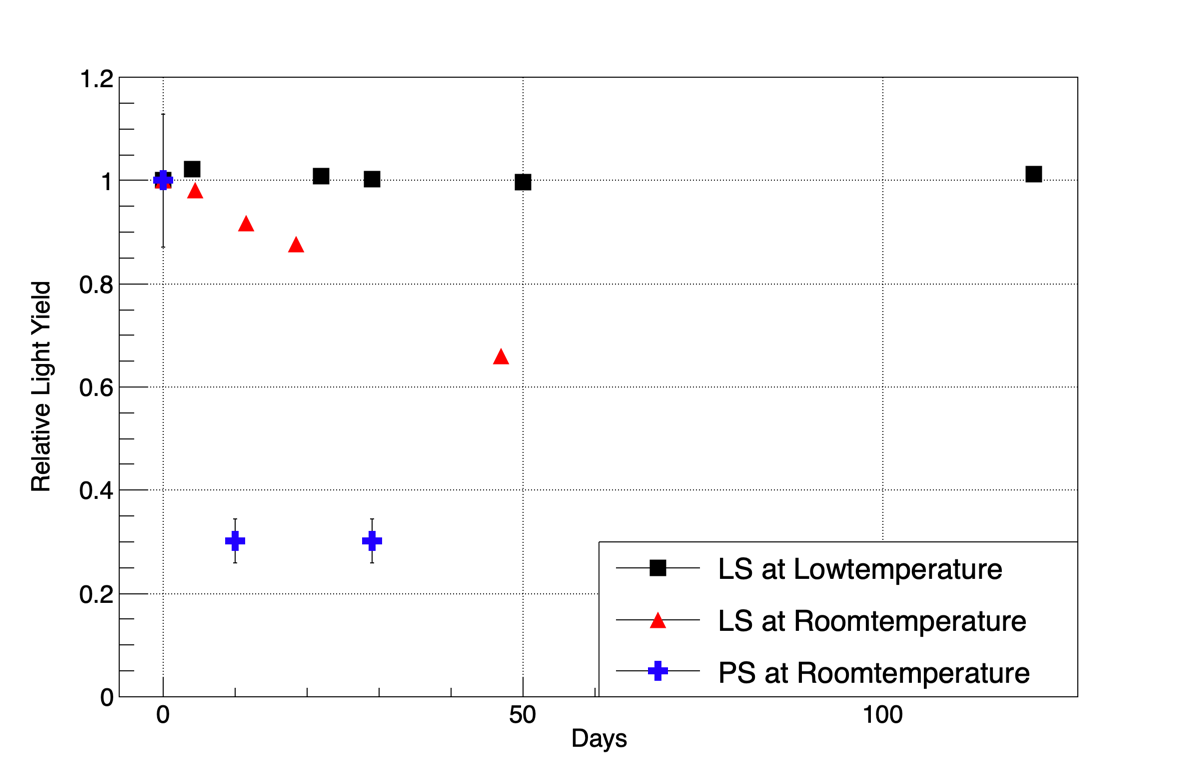}
  \end{center}
  \caption{Single photoelectron relative outcomes over meansurement.
  }
  \label{spe}
\end{figure*}

\section{Results}
NaI(Tl) crystal has been encapsulated by organic scintillators to prevent humidity and to tag external radiation.
LAB-based liquid and plastic scintillator are used to encompass the crystal with direct optical contact.
Light yields of NaI(Tl) are measured to be (11.3~$\pm$~1.0) and (1.2~$\pm$~0.2)~pe/keV for
liquid and plastic scintillator encapsulation, respectively.
PSD analysis confirms that organic and inorganic scintillator light signals are relatively well separated.
(7.7~$\sigma$ at 400 keV).
Furthermore, pile-up signals of those two different waveforms are identified.
When any radioactive isotope decay events occuring near the boundary between crystal and organic scintillator,
it also could be identified by PSD analysis.
We have investigated long-term light yield stability at low temperature (-35 $^{\circ}$C) for liquid scintillator encapsulation.
This study shows that a stable light yield has been achieved over four months, since
humidity effects on the crystal surface might have been reduced below freezing temperature.
In contrast, ambient temperature measurement shows approximately 30\% light yield reduction over one month.

\section{Conclusion}

 The NaI(Tl) scintillation crystal has been encapsulated with liquid and plastic scintillators, in order to prevent direct air contact,
and used propoesd setup to investigate the surface alpha decay events,
which could cause serious radioactive background for dark matter searches.
We show that liquid scintillator is suitable
to protect NaI(Tl) crystal from the moisture under optimized conditions(-35 $^{\circ}$C).
Events caused by organic and NaI(Tl) scintillator were separated
using pulse shape discrimination analysis.
We will study the surface $^{210}$Pb($^{210}$Po) contamination, which could
be identifiable by tagging $^{210}$Po alpha particles in an organic scintillator
and measuring nuclear recoil ($^{206}$Pb) in the crystal with the similar setup
as discussed in the current paper.

\section{Acknowledgments}
This research was funded by the Institute for Basic Science (Korea) under project code IBS-R016-A1; NRF (National Research Foundation of Korea) Grant funded by the Korean Government (NRF-2016-Fostering Core Leaders of the Future Basic Science Program/Global Ph.D. Fellowship Program.); National Research Foundation of Korea (NRF) funded by the Ministry of Science and Technology, Korea (MEST, No. 2018R1A6A1A06024970).

\end{document}